\documentclass[12pt]{article}
\usepackage{amsmath,amssymb,amsfonts,epsfig,graphicx,euscript}

\newcommand{\bse}{\begin{subequations}}
\newcommand{\ese}{\end{subequations}}
\newcommand{\be}{\begin{equation}}
\newcommand{\ee}{\end{equation}}
\newcommand{\bea}{\begin{eqnarray}}
\newcommand{\eea}{\end{eqnarray}}
\newcommand{\ba}{\begin{array}}
\newcommand{\ea}{\end{array}}

\newcommand{\ie}{{\it i.e.}}

\begin{document}
\hfill%
\vbox{
    \halign{#\hfil        \cr
           IPM/P-2010/032\cr
                     }
      }
\vspace{1cm}
\begin{center}
{ \Large{\textbf{Conductivity at finite 't Hooft coupling from
AdS/CFT}
\\}}
\vspace*{2cm}
{\bf M. Ali-Akbari$^1$ and K. Bitaghsir Fadafan$^2$}\\%
\vspace*{0.4cm}
{\it {$^1$School of Physics, Institute for Research in Fundamental Sciences (IPM)\\
P.O.Box 19395-5531, Tehran, Iran}}  \\
{E-mails: {\tt aliakbari@theory.ipm.ac.ir}}\\%
{\it {$^2$Physics Department, Shahrood University of Technology,\\
P.O.Box 3619995161, Shahrood, Iran }}\\
{E-mails: {\tt bitaghsir@shahroodut.ac.ir}}%
\vspace*{1.5cm}
\end{center}

\vspace{.5cm}
\bigskip
\begin{center}
\textbf{Abstract}
\end{center}
We use the AdS/CFT correspondence to study the DC conductivity of
massive $\mathcal{N} = 2$ hypermultiplet fields in an $\mathcal{N} =
4\, SU(N_c)$ super-Yang-Mills theory plasma in the large $N_c$ and
finite 't Hooft coupling. We also discuss general curvature-squared
and Gauss-Bonnet corrections on the DC conductivity.

\newpage
\tableofcontents

\section{Introduction}
Anti-de Sitter/conformal field theory correspondence ($AdS/CFT$)
conjectures that IIB string theory on $AdS_5\times S^5$ background
is dual to $D=4\ {\cal{N}}=4\ SU(N_c)$ super Yang-Mills theory (SYM)
\cite{Maldacena:1997re}. In the limits of large $N_c$ and large 't
Hooft coupling $\lambda=g_{YM}N_c^2$, the SYM theory is dual to IIb
supergravity which is low energy effective theory of superstring
theory. As a result a thermal SYM theory corresponds to supergravity
in an AdS-Schwarzschild background where the SYM theory temperature
is identified with the Hawking temperature of the $AdS$ black hole
\cite{Witten:1998zw}.

$AdS/CFT$ idea has been applied to study different aspects of
strongly coupled SYM theory. Recently the application of this
duality in condensed matter physics (called $AdS/CMT$) has been
studied \cite{Review} . This duality is very useful to study certain
strongly coupled systems in $CMT$ by holography techniques and to
understand better their properties. A quantity attracting attentions
and computing is conductivity \cite{Karch}. In order to compute
conductivity, it is necessary to add spinor fields in fundamental
representation. As it is well-known fields in the $\mathcal{N}=4$
SYM theory are in adjoint representation and by introducing
$\mathcal{N}=2$ hypermultiplets one can include fields in the
fundamental representation. To do so on gravity side, $N_f$ flavor
branes are introduced in the probe limit meaning we have only
$N_f\ll N_c$ of them and the $AdS$ background is left unchanged. In
other words, open string degrees of freedom are now considered. On
the gauge theory side this corresponds to ${\cal{N}}=2$
supersymmetric hypermultiplet flavor fields propagating in SYM
theory.  The local $U(N_f)$ symmetry on the D7-branes corresponds to
global $U(N_f)$ symmetry whose $U(1)_B$ subgroup may be identified
with baryon number. Non-dynamical electric and magnetic fields can
be coupled to $U(1)_B$ charge and we then expect a constant, nonzero
currents.
Hence the conductivity tensor $\sigma_{ij}$ is identified by %
\be %
 <J_i>=\sigma_{ij}E_j\,.
\ee %
Electric and magnetic fields produce diagonal and off-diagonal
elements in conductivity tensor respectively. In fact on the gravity
side currents, electric and magnetic fields are introduced as
nontrivial gauge
fields living on the D7-branes.%

The DC conductivity in "strange metals" has been studied in
\cite{strangeM,Fadafan:2009an}. It was found that conductivity is
independent of the dimensionality of a probe brane in the background
created by a stack of D$_p$-branes \cite{Karch:2009eb}. Also it was
shown that the properties of the DC conductivity can be studied by
that of the single heavy quark. One should consider massive carriers
as endpoint of strings which end on flavor branes and use the
quasi-particle description. Then it would be straightforward to
calculate the DC conductivity from the drag force
calculations\cite{Fadafan:2009an,Charmousis:2010zz}. We discuss this
approach in the last section.%

In this paper we study higher derivative corrections on the DC
conductivity. These corrections on the gravity side correspond to
finite coupling corrections on the gauge theory side. The main
motivation to consider corrections comes from the fact that string
theory contains higher derivative corrections arising from stringy
effects. In the case of $\mathcal{N}=4$ SYM theory, the leading
order correction in $1/\lambda$ arises from stringy correction to
the low energy effective action of type $\amalg$b supergravity,
$\alpha'^3 {\cal{R}}^4$. We study ${\cal{R}}^4$ and ${\cal{R}}^2$
corrections to the DC conductivity. An understanding of how these
computations are affected by finite $\lambda$ corrections may be
essential for
more precise theoretical predictions.%

The article is organized as follows. In the next section, we will
give a brief review of \cite{Karch,OBannon} and then follow the same
direction and study Ohmic and Hall conductivities in the presence of
corrections in section 3. Moreover in section 4, we consider general
${\cal{R}}^2$ and Gauss-Bonnet corrections and in the
last section we draw our conclusions and summarize our results. %

\section{Review of conductivity}
In this section we will give a brief review of \cite{Karch,OBannon}.
The AdS-Schwarzchild metric, in units where the radius of $AdS$ is
one, is \footnote{A change of variable as $uz=1+\frac{z_h^4}{z^4}$
relates our coordinate to what is used in \cite{Karch}.}
\be\label{metricform}%
 ds^2=G_{tt}\ dt^2+G_{xx}d\vec{x}^2+G_{uu}\ du^2+d^2\Omega_5\,. %
\ee %
Here $d\vec{x}^2$ is the three dimensional space and the metric
functions are given by%
\bse\begin{align}
 G_{tt}&=-u^2(1-\frac{u_0^4}{u^4}),\\
 \label{GXX}G_{xx}&=u^2,\\
 G_{uu}&=u^{-2}(1-\frac{u_0^4}{u^4})^{-1},
\end{align}\ese %
where $u_0$ is event horizon. Also $S^5$ metric is%
\be %
d\Omega_5^2=d\theta^2+sin^2\theta\,d\psi^2+cos^2\theta^2\,d\Omega_3^2\,.
\ee %
where $d\Omega_3^2$ is the metric of $S^3$ and $\theta$ runs from
zero to $\frac{\pi}{2}$. In addition the Hawking temperature is given by %
\be\label{temprature} %
 T=\frac{u_0}{\pi}\,.
\ee %

In order to find conductivity a number $N_f$ of D7-branes filling
$AdS_5$ and wrapping the $S^3\subset S^5$ have been considered in
the AdS-Schwarzchild background. Moreover worldvolume gauge fields
$A_t(u)$, $A_x(u,t)=-Et+f_x(u)$ and $A_y(u,t)=Bx+f_y(u)$ as well as
the $\theta(u)$ describing the position $S^3$ on the $S^5$ are
turned on. The above system is described by Dirac-Born-Infeld
action. From the gauge field equations of motion, conserved charges
$Q$, $I^x$ and $I^y$ associated with $A_t(u)$, $A_x(u,t)$ and
$A_y(u,t)$ are found. The conserved charges which are sub-leading,
normalizable terms of asymptotic expansion can be related to
expectation value of dual operators \ie
\be %
 <J^t>=Q,\ \ \ \ <J^x>=I^x,\ \ \ \ <J^y>=I^y. %
\ee %
Having worked out the gauge fields in terms of $Q$ and $B$ one can
write the on-shell action as follows \cite{OBannon}%
\be\label{onshellaction} %
 S_{D7}=-{\cal{N}}^2(2\pi\alpha')^2\int\,d^4x
 cos^6\theta\,g_{xx}^2\,\sqrt{g_{tt}g_{uu}}\frac{\xi}{\sqrt{\xi\chi-a^2}},
\ee%
where ${\cal{N}}=\frac{\lambda}{(2\pi^4)}N_fN_c$. $g_{tt}(=|
g_{tt}|),\ g_{uu}$ and $g_{xx}$ are components of induced metric and
\bse\label{conductivity}\begin{align}%
 \label{1}\xi&=g_{tt}g_{xx}^2+(2\pi\alpha')^2(g_{tt}B^2-g_{xx}E^2),\\
 \label{2}\chi&=g_{tt}g_{xx}^2\left({\cal{N}}^2(2\pi\alpha')^4g_{xx}\cos^6\theta\right)
 +(2\pi\alpha')^2\left(g_{tt}Q^2-g_{xx}(I_x^2+I_y^2)\right),\\
 \label{3}a&=(2\pi\alpha')^2\left(g_{tt}I^xB+g_{xx}I^yE\right).
\end{align}\ese %
$\xi$, as a function of $u$, has negative and positive value at the
horizon and near the boundary respectively. In fact at a specific
point, say $u_*$, we have $\xi(u_*)=0$. Hence $u_*$ is %
\be %
u_*^4=u_0^4{\cal{F}}(e,b)=\pi^4T^4{\cal{F}}(e,b)= g_{xx}^2(u_*)
\ee %
where\footnote{Notice that $\alpha'^{-2}=4\pi g_sN_c=g_{YM}^2N_c=\lambda$.} %
\be\begin{split} %
 {\cal{F}}(e,b)&=\frac{1}{2}\left(1+e^2-b^2+
 \sqrt{(1+e^2-b^2)^2+4 b^2}\right)\cr
 e&=\frac{E}{\frac{\pi}{2}\sqrt{\lambda}T^2},\ \
 b=\frac{B}{\frac{\pi}{2}\sqrt{\lambda}T^2},\ \
 q=\frac{Q}{\frac{\pi}{2}\sqrt{\lambda}T^2}
\end{split}\ee %
Reality condition of $\sqrt{\xi\chi-a^2}$ imposes that two functions
$\chi$ and $a$ must vanish at $u=u_*$. By setting \eqref{2} and
\eqref{3} to zero at $u_*$ and converting to field theory
quantities, Ohmic and Hall conductivities become %
\bse\begin{align} %
 \sigma_{xx}&=\sqrt{\frac{N_f^2N_c^2\cos^6\theta_*}{16\pi^2}\frac{{\cal{F}}^{3/2}}{b^2+{\cal{F}}}T^2
 +\frac{q^2{\cal{F}}}{(b^2+{\cal{F}})^2}}\\
 \sigma_{xy}&=\frac{q b}{b^2+{\cal{F}}}
\end{align}\ese %

The Ohmic conductivity depends on magnetic field, electric field and
the temperature of the field theory, simplicity we name it as
$\sigma(B,E,T)$. This quantity has two main terms%
\be
\sigma_{xx}(B,E,T)=\sqrt{\hat{\sigma}_0^2+\hat{\sigma}^2}\label{sigma2}
 \ee%
where $\hat{\sigma}_0$ arises from thermally produced pairs of
charge carriers. By increasing the mass of carriers,
$\hat{\sigma}_0$ can be made arbitrary small and the leading term in
conductivity will be $\hat{\sigma}$. This term can be found by
studying the properties of a moving single string and calculating
drag force \cite{Karch}. We will discuss this point in the
discussion section.

In next section we will follow the same direction and study above
conductivities in the presence of correction.

\section{${\cal{R}}^4$ corrections to AdS-Schwarzschild black
brane}

Since $AdS/CFT$ correspondence refers to complete string theory, one
should consider the string corrections to the 10D supergravity
action. The first correction occurs at order $(\alpha')^3$
\cite{alpha2}. In the extremal $AdS_5\times S^5$ it is clear that
the metric does not change \cite{Banks}, conversely this is no
longer true in the non-extremal case. Corrections in inverse 't
Hooft coupling $1/\lambda$ which correspond to $\alpha^{\prime}$
corrections on the string theory side were found in \cite{alpha2}.
Higher derivative corrections {\em{i.e}}. ${\cal{R}}^2$ and
${\cal{R}}^4$ on the rotating quark-antiquark system in the hot
plasma and on the drag force on a moving heavy quark have been
investigated in \cite{AliAkbari:2009pf, Fadafan,Fadafan:2008gb}.

Functions of the $\alpha^{\prime}$-corrected metric are given by
\cite{alpha1}
\begin{eqnarray}\label{correctedmetric}
G_{tt}&=&-u^2(1-w^{-4})T(w),\nonumber\\
G_{xx}&=&u^2 X(w),\nonumber\\
G_{uu}&=&u^{-2}(1-w^{-4})^{-1} U(w),
\end{eqnarray}
where%
\begin{eqnarray}
T(w)&=&1-k\bigg(75w^{-4}+\frac{1225}{16}w^{-8}-\frac{695}{16}w^{-12}\bigg)+\dots ,\nonumber\\
X(w)&=&1-\frac{25k}{16}w^{-8}(1+w^{-4})+\dots,\nonumber\\
U(w)&=&1+k\bigg(75w^{-4}+\frac{1175}{16}w^{-8}-\frac{4585}{16}w^{-12}\bigg)+\dots,\
\end{eqnarray}
and $w=\frac{u}{u_0}$. There is an event horizon at $u=u_0$ and the
geometry is asymptotically $AdS$ at large $u$ with a radius of
curvature $R=1$. The expansion parameter $k$ can be expressed in
terms
of the inverse 't Hooft coupling as %
\be\label{k}%
 k=\frac{\zeta(3)}{8}\lambda^{-3/2}\sim 0.15\lambda^{-3/2}.
\ee %
The temperature is given by%
\be %
 T_{{\cal{R}}^4}=\frac{u_0}{\pi(1-k)}.
\ee %
In order to find conductivities we set all of equations
\eqref{conductivity} to zero where the induced metric components
will now be computed by \eqref{correctedmetric}.

\subsection{Ohmic conductivity}

Here we focus on a case where the magnetic field has been turned off
and \eqref{1}, after setting to zero, becomes %
\be\begin{split}\label{kconduc} %
 w^{16}-(1+75k+e^2)w^{12}-k\left(\frac{25}{8}w^8-120\,w^4+\frac{335}{8}\right)=0,
\end{split}\ee %
up to first order in $k$ perturbatively. $u_*=u_0w_*$ is a real
parameter and takes value between $u_0$ and infinity. These two
conditions tell us that just a root of \eqref{kconduc} is acceptable
for $w_*$ among sixteen roots. Now \eqref{2} and \eqref{3} must be
set to zero. \eqref{3} leads to
$I^y=0$ and \eqref{2} becomes %
\be\label{Ohmic} %
 \sigma_{xx}=\frac{I^x}{E}=\sqrt{{\cal{N}}^2(2\pi
 \alpha')^4g_{xx}\cos^6\theta_*+\frac{(2\pi
 \alpha')^2}{g_{xx}^2}Q^2}\,.
\ee %
where $g_{xx}$ evaluated at $u_*$, which is found by solving \eqref{kconduc}, is given by %
\be\label{gxx1} %
 g_{xx}(u_*)=\pi^2T^2\left(\sqrt{1+e^2}+\frac{5k\left(-10+e^2 \left(163+120
 e^2 \left(3+e^2\right)\right)\right)}
 {16 \left(1+e^2\right)^{7/2}}\right),
\ee %
Finally by substituting \eqref{gxx1} in \eqref{Ohmic} the Ohmic
conductivity becomes %

\be\begin{split} %
 \sigma^{\mathcal{R}^4}_{xx}&=\sigma^{\mathcal{N}=4}_{xx}+\cr
 & \frac{5\,\zeta(3)\,\lambda^{-3/2} }{256\,\left(1+e^2\right)^4}\,
 \bigg(-10+e^2\left(163+120 e^2 \left(3+e^2\right)\right)\bigg)
 \,\bigg(\sigma^{\mathcal{N}=4}_{xx}-\frac{3\frac{q^2}{1+e^2}}{\sigma^{\mathcal{N}=4}_{xx}}\bigg)\,,
\end{split}\ee %
where%
\be\label{conductivity1} %
 \sigma^{{\mathcal{N}=4}}_{xx}=\sqrt{\frac{N_f^2N_c^2T^2}{16\pi^2}\sqrt{1+e^2}\cos^6\theta_*+\frac{q^2}{1+e^2}}\,.
\ee %

We would like to emphasize two points about our result. First, at
finite temperature for a large value of electric field second term
in the conductivity equation coming from correction goes to zero
which means that correction of metric is negligible. Second,
depending on other parameters, conductivity increases, decreases or
does not change. In particular by taking large mass limit
$\cos\theta_*\rightarrow 0$ [1, 2] there is a critical value of
electric field $e_c = 0.233745$ where the correction effect
vanishes. As a result conductivity increases when $e < e_c$ and vice
versa. In the limit of zero density, if $e > e_c$ or $e < e_c$ we
have a positive or negative correction effect, respectively.

\subsection{Conductivities in the presence of magnetic field}

In this section, we study the effect of 't Hooft coupling correction
on the conductivities when the magnetic field is turned on. In order
to do that by setting \eqref{conductivity} to zero, we have %
\bse\label{hall}\begin{align}%
 \label{hall1}w(w&-1)(16w^3-1200kw^2-1275kw+645k)+e^2w(-16w^3
 +25kw+25k)\\ \nonumber
 &+b^2(w-1)(16w^3-1200kw^2-1225kw+695k)=0,\\
 \label{hall2}\sigma_{xx}&=\frac{g_{xx}}{g_{xx}^2
 +(2\pi\alpha')^2B^2}\sqrt{(2\pi\alpha')^4{\cal{N}}(g_{xx}^2+(2\pi\alpha')^2B^2)
 g_{xx}\cos^6\theta_*+(2\pi\alpha')^2Q^2},\\ %
 \label{hall3}\sigma_{xy}&=\frac{(2\pi\alpha')^2QB}{g_{xx}^2+(2\pi\alpha')^2B^2}\,.
\end{align}\ese %
where as before $g_{xx}$ and $\cos\theta$ in \eqref{hall2} and
\eqref{hall3} are evaluated at $w=w_*$. In this case $w_*$ will be
found by solving \eqref{hall1}. However this equation is complicated
one can solve it numerically. We assume different values for
parameters and discuss behavior of Ohmic and Hall conductivities in
terms of them. In the massive case, $\cos\theta\rightarrow 0$, we
have plotted conductivities versus $e$ and $b$ in Fig. 1-4. Our
figures have been plotted at fixed temperature($u_0=3$) and for
three different values of $k$(=0.0001,0.001,0.01).

Ohmic conductivity has been plotted in Fig. 1. In the left plot of
this figure $b=10$. Notice that from (\ref{k}), increasing $k$
corresponds to decreasing 't Hooft coupling constant $\lambda$. For
small values of $\lambda$, there is a specific maximum value for
Ohmic conductivity. It is clearly seen that by increasing 't Hooft
coupling constant the shape of Ohmic conductivity changes and it
approximately starts from its maximum value. In the right plot of
Fig. 1 $b=30$. As one finds from (\ref{hall}), by increasing $b$ the
value of Ohmic conductivity will be decreased. This behavior of
Ohmic conductivity is clearly shown in this plot. In addition by
increasing $b$, the maximum value of Ohmic conductivity appears at
larger value of $e$.

\begin{figure}[ht]
\centerline{\includegraphics[width=3in]{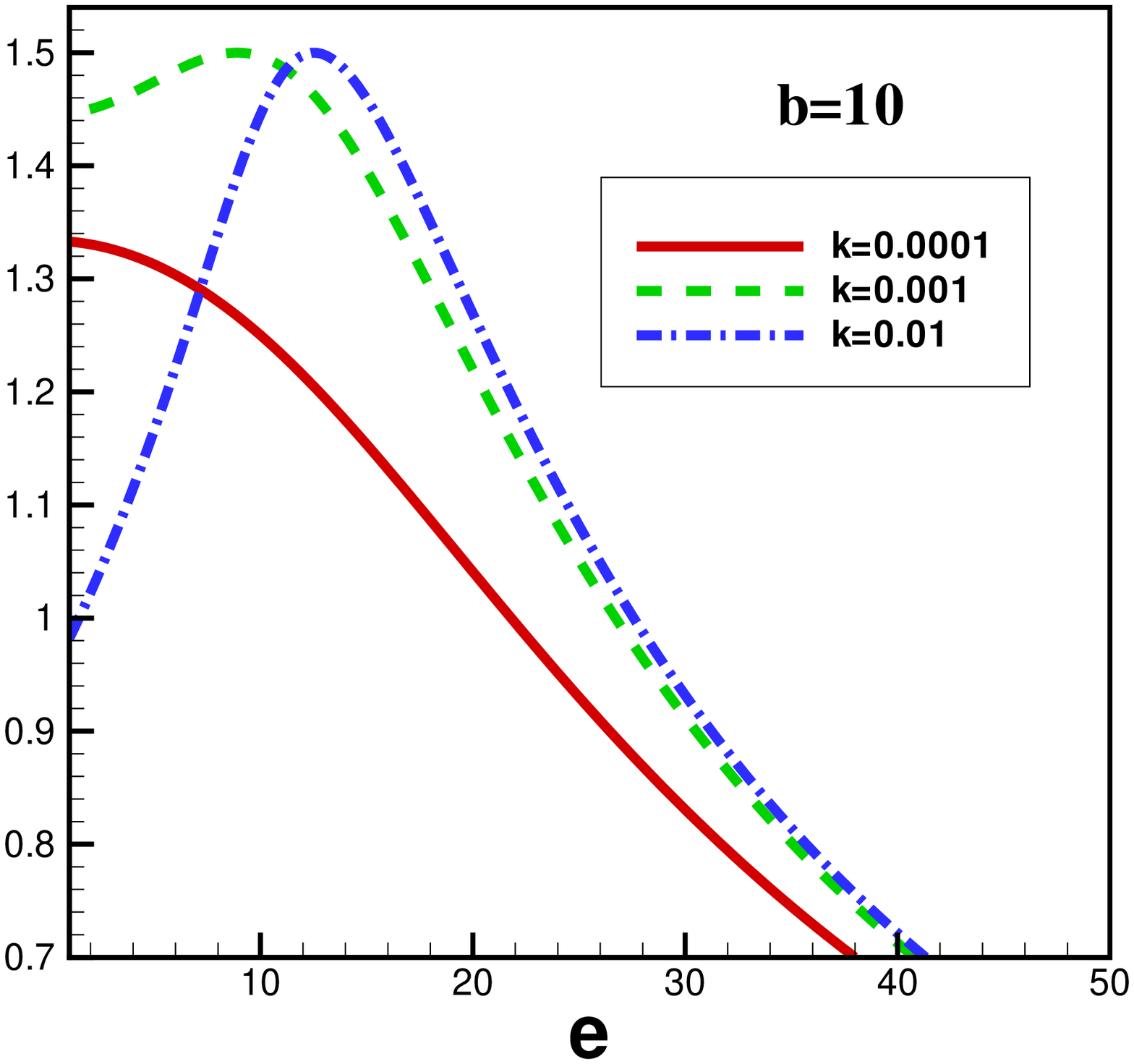},\includegraphics[width=3in]{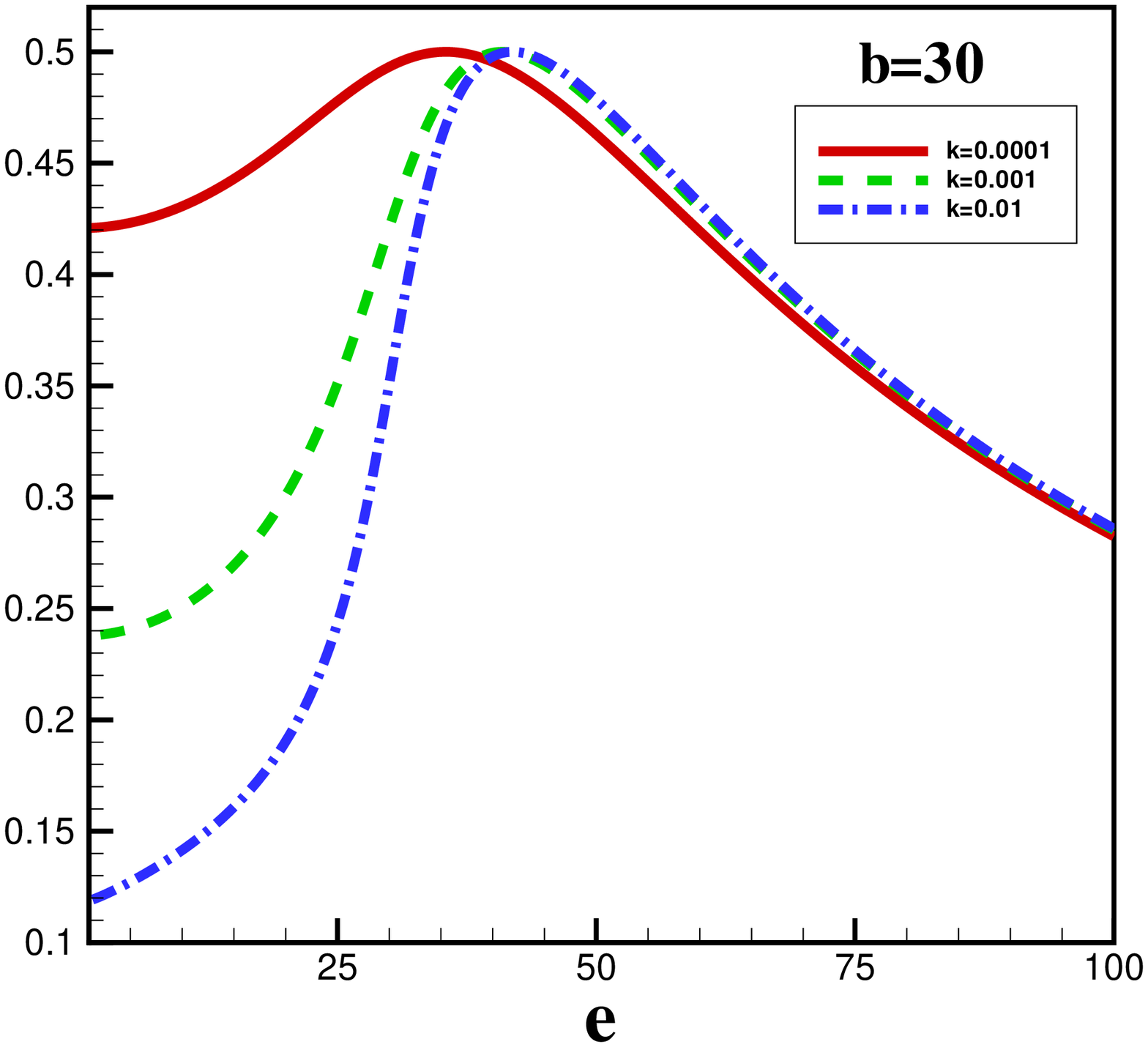}}%
\caption{$\sigma_{xx}$ versus $e$ for different values of $k$ at
fixed $b$. Left: $b=10$. Right: $b=30$. }
\end{figure}%

It would also be interesting to investigate behavior of Ohmic
conductivity in terms of $b$ at constant $e$. In the left and right
plots of Fig 2, we have plotted Ohmic conductivity for $e=10$ and
$e=20$, respectively.  As it is clear from these plots Ohmic
conductivity starts from a maximum value and by increasing $b$ it
decreases. The main effect of increasing 't Hooft coupling constant
is to decrease Ohmic conductivity value. The right plot of Fig 2
shows the same results. The point is that increasing the value of
$e$ causes a decreasing of Ohmic conductivity value.

\begin{figure}[ht]
\centerline{\includegraphics[width=3in]{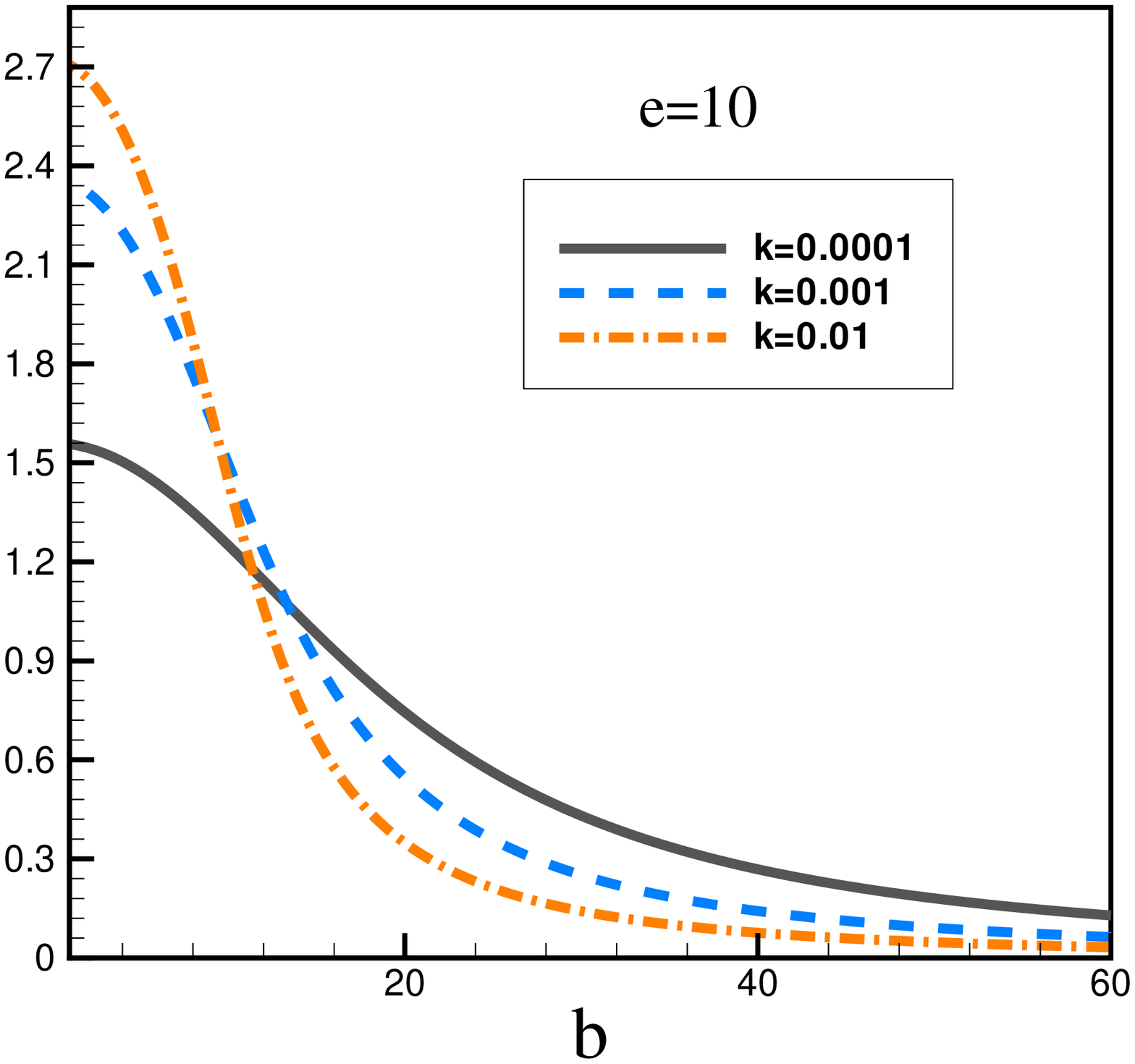},\includegraphics[width=3in]{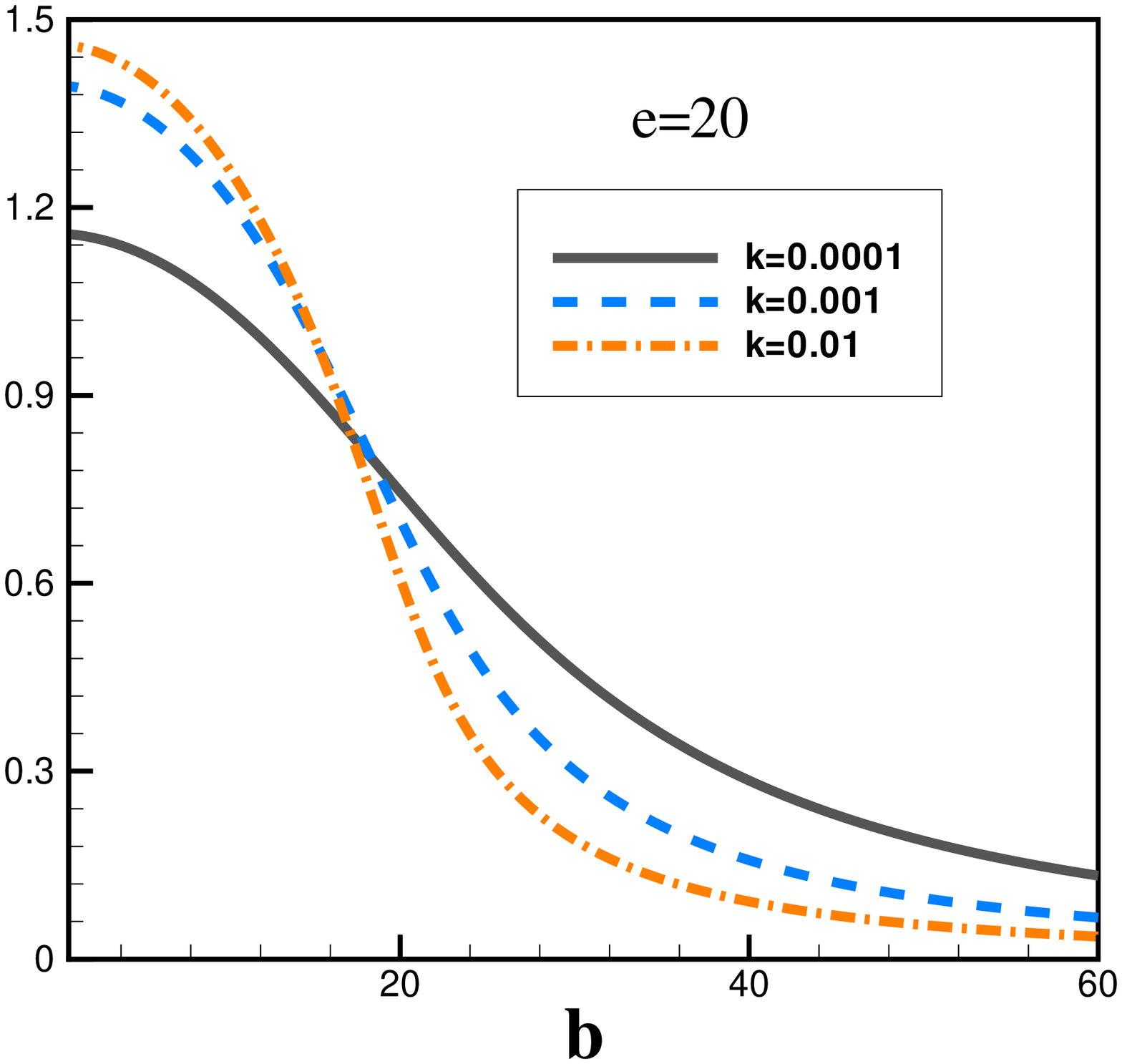}}
 \caption{$\sigma_{xx}$ versus $b$ for different values of $k$ at
fixed $e$. Left: $e=10$. Right: $e=20$.}
\end{figure}%

In following two paragraphs we study Hall conductivity versus $e$
and $b$. Hall conductivity in terms of $e$ is plotted in Fig. 3 at
constant $b$. In the left plot of this figure $b=10$. It is clearly
seen that Hall conductivity also starts from its maximum value and
it decreases by increasing $e$. Moreover if 't Hooft coupling
constant increases, the maximum value of Hall conductivity will
decrease. Comparing with the left plot of Fig. 3, in the right plot
Hall conductivity has a smaller value. This fact, which Hall
conductivity decreases by increasing $b$, is consistent with
\eqref{hall3}.

\begin{figure}[ht]
\centerline{\includegraphics[width=3in]{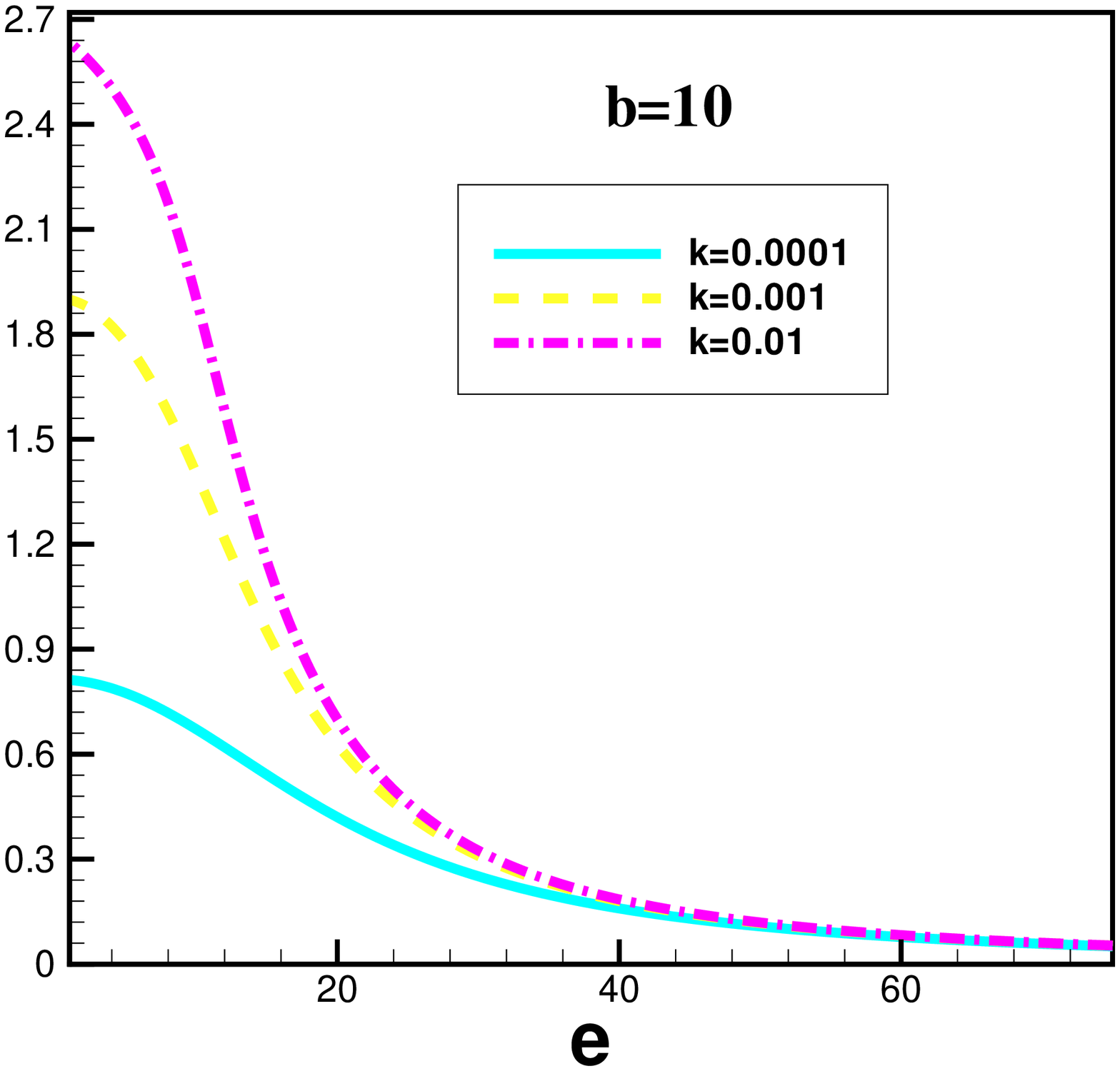},\includegraphics[width=3in]{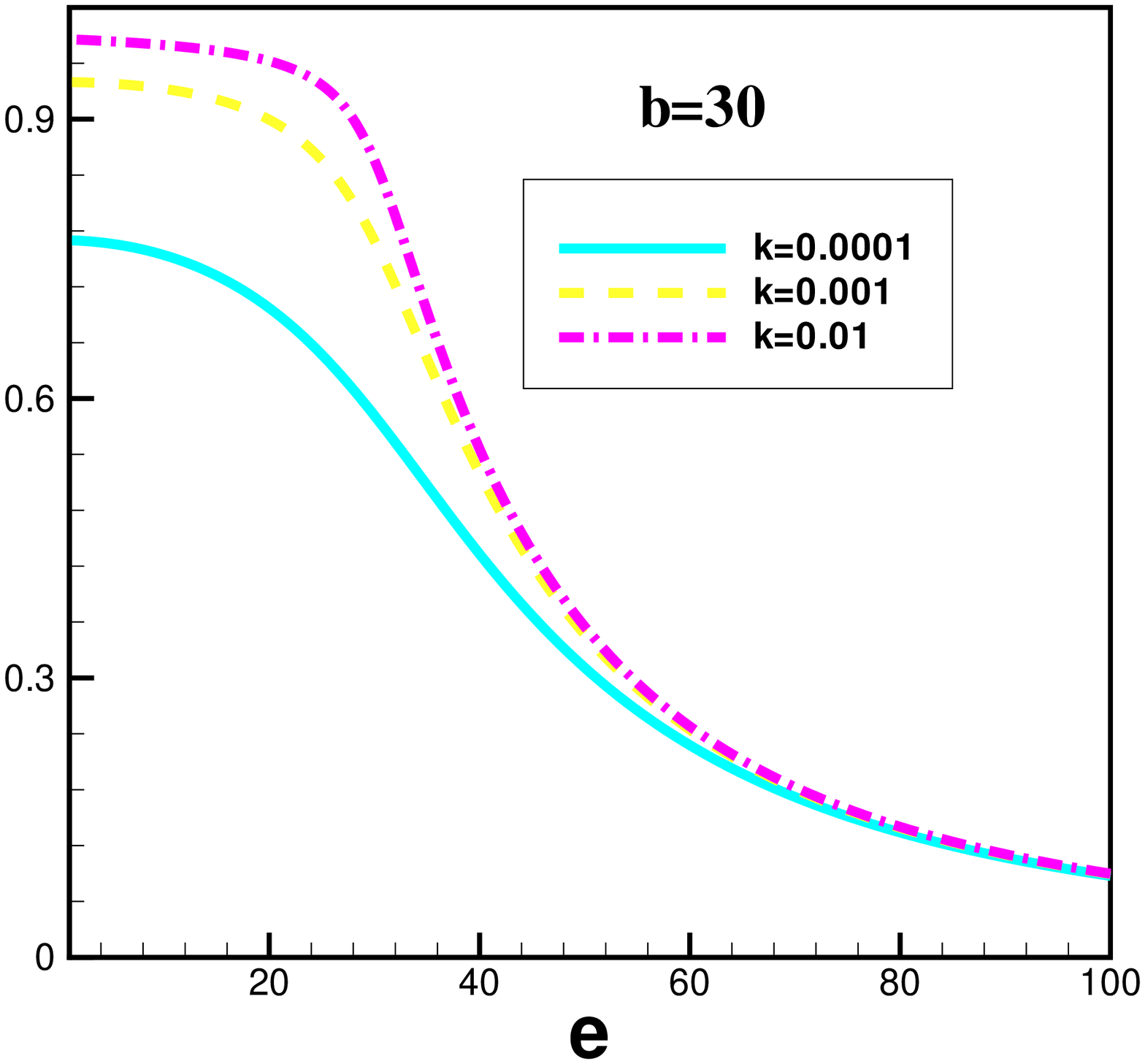}
} \caption{$\sigma_{xy}$ versus $e$ for different values of $k$ at
fixed  $b$. Left: $b=10$. Right: $b=30$. }
\end{figure}%

\begin{figure}[ht]
\centerline{\includegraphics[width=3in]{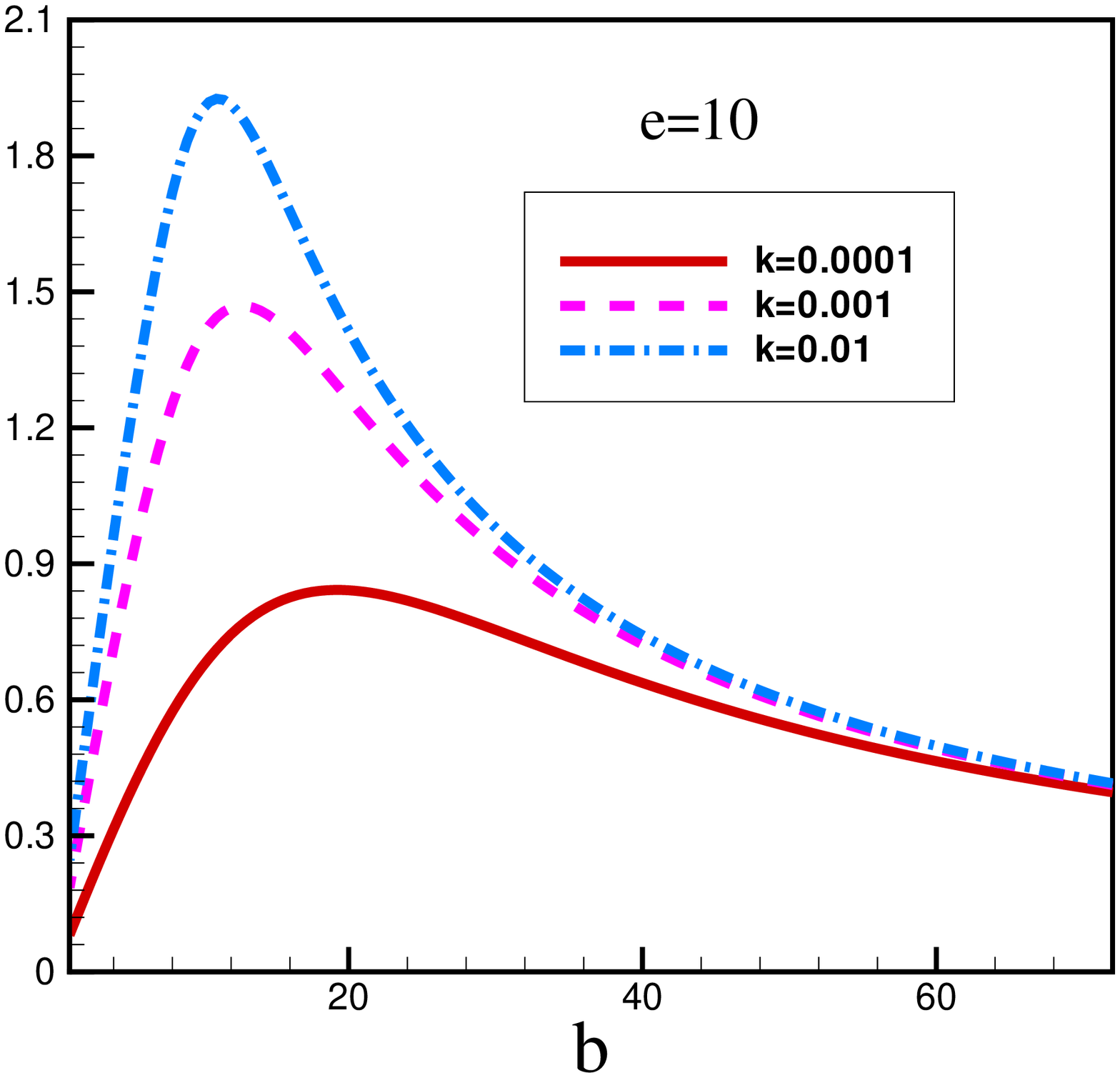},\includegraphics[width=3in]{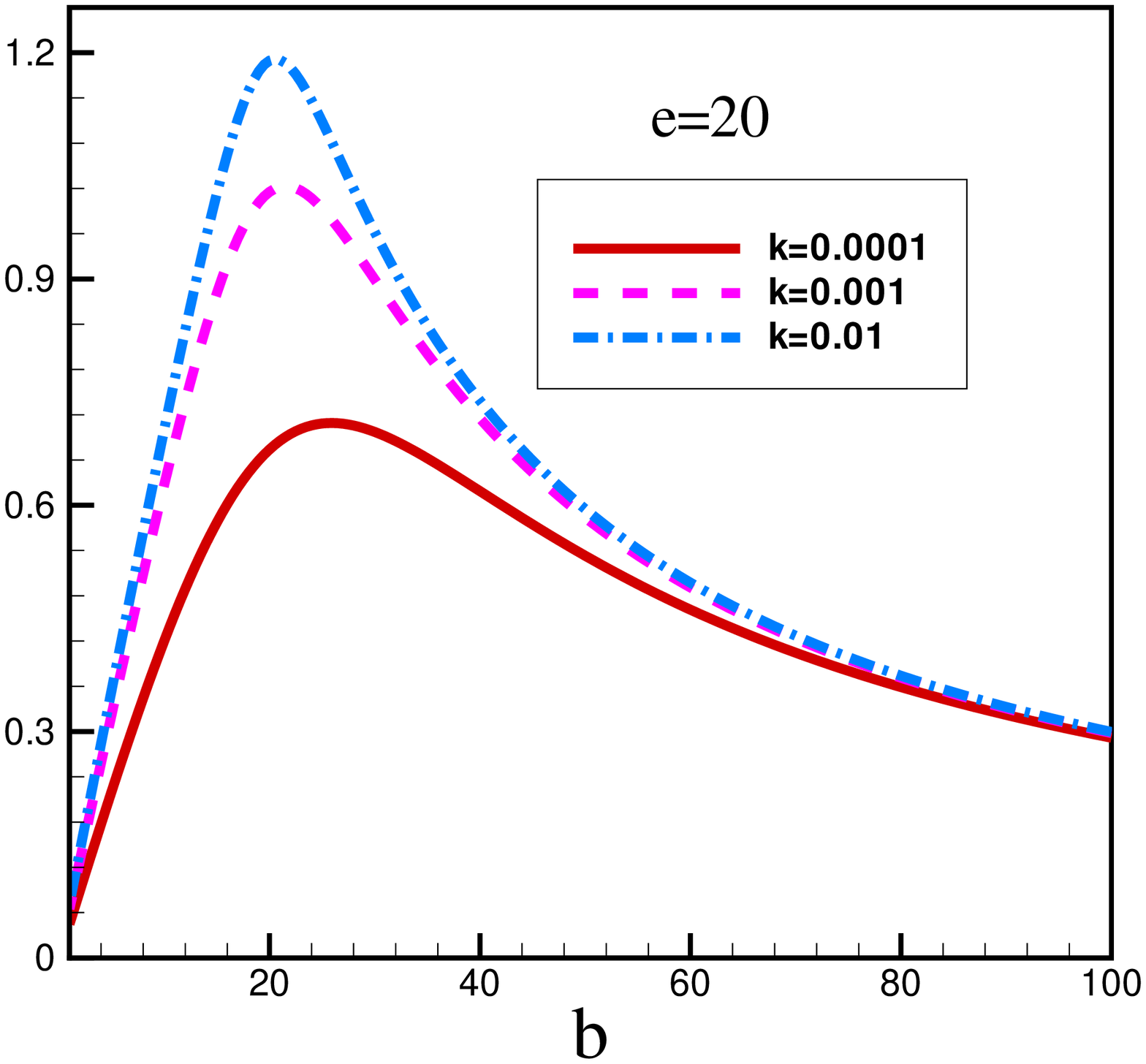}}
\caption{$\sigma_{xy}$ versus $b$ for different values of $k$ at
fixed  $e$. Left: $e=10$. Right: $e=20$.}
\end{figure}%
Hall conductivity versus $b$ is plotted in Fig. 4 at constant $e$.
Two cases, which are $e=10$ and $e=20$, are considered. As it is
obviously seen, similar to Fig. 2 and 3, there is a maximum value
for Hall conductivity. By increasing 't Hooft coupling constant this
maximum value decreases and in fact this is the most important
effect of 't Hooft coupling correction on the Hall conductivity. The
right plot of Fig. 4 indicates that Hall conductivity has a smaller
value when $e=20$.

\section{${\cal{R}}^2$ corrections to AdS-Schwarzschild black brane}

We consider the curvature-squared corrections on the AdS black brane
solution  and study DC conductivity in this background.
${\cal{R}}^2$ corrections have been introduced in
\cite{Blau:1999vz,Fayyazuddin:1998fb}. It is important to point out
that the first higher derivative correction in weakly curved type
IIb backgrounds enters at order $\mathcal{R}^4$, and not
$\mathcal{R}^2$, so we will not predict effect of curvature-squared
corrections on the $\mathcal{N}=4$ SU(N) SYM. At first we consider
the general $\mathcal{R}^2$ corrections in \eqref{R2} and calculate
Ohmic conductivity and then extend the calculation to the case of
Gauss-Bonnet (GB) gravity. Fortunately, we find analytic results in
both cases. Hall conductivity in the presence of
curvature-squared corrections can be studied, too. Here we only discuss Ohmic conductivity.%

The black brane solution of $AdS_5$ space with the curvature-squared
corrections is \cite{Kats:2007mq}
\begin{eqnarray}\label{R2}
ds^2=- \kappa(\frac{u^2}{R^2})f(u)
dt^2+(\frac{u^2}{R^2})d\vec{x}^2+\frac{1}{(\frac{u^2}{R^2})f(u)}du^2,
\end{eqnarray}
where
\begin{equation}
 f(u)=1-\frac{u_0^4}{u^4}+\alpha+\gamma \frac{u_0^8}{u^8}.
\end{equation}
and $\alpha$ and $\gamma$ are small variables. The scaling factor
$\kappa=\frac{1}{1+\alpha}$ was included for the speed of light in
the boundary to be unity.  Also $u$ denotes the radial coordinate of
the black brane geometry and $t, \vec{x}$ label the directions along
the boundary at the spatial infinity. In these coordinates the event
horizon is located at $f(u_0)=0$ where $u_0$ can be found by solving
this equation. The boundary is located at infinity and the geometry
will be asymptotically AdS with a radius of curvature $R=1$. The
temperature is given by
\begin{equation}
T_{\mathcal{R}^2}=\frac{u_0}{\pi} \left( 1+ \frac{1}{4} \alpha -
\frac{5}{4} \gamma\right)\label{TR2}.
\end{equation}
As what it was done in section 3.1, \eqref{1} after setting to zero,
becomes
\be\label{r2equation} %
 w\left((1+\alpha)w+\frac{e^2}{\kappa}(1+\frac{\alpha}{4}-\frac{5\gamma}{4})^4-1\right)+\gamma=0,
\ee %
where %
\be %
 e=\frac{E}{\frac{\pi}{2}\sqrt{\lambda}\,T_{\mathcal{R}^2}^2}.
\ee %
Obviously by setting $\alpha$ and $\gamma$ to zero one must find
Ohmic conductivity in AdS-Schwarzschild background brought in
\eqref{conductivity1}. This virtue selects an acceptable root of
\eqref{r2equation} which was already called $w_*$. It is then
straightforward to find Ohmic conductivity from \eqref{Ohmic} that
is given by %
\be\begin{split}\label{R2cond} %
 (\sigma^{\mathcal{R}^2}_{xx})^2&=\frac{512\,(1+\alpha)q^2}{256+(1+\alpha)\left( 4+\alpha-5\gamma \right)^4\,e^2+
 \sqrt{M\left(\alpha,\gamma,e\right)}}\cr
 &+\frac{1}{(16\pi)^2\sqrt{2}}
 \frac{N_c^2N_f^2T^2_{\mathcal{R}^2}}{(1+\frac{\alpha}{4}-\frac{5}{4}\gamma)^2}\cos^6{\theta_*}\cr
 &\times \left(\frac{256+(1+\alpha)\left( 4+\alpha-5\gamma \right)^4\,e^2+
 \sqrt{M\left(\alpha,\gamma,e\right)}}{1+\alpha}\right)^{\frac{1}{2}},
\end{split}\ee %
where%
\be M\left(\alpha,\gamma,e\right)=-2^{18}\,\gamma(1+\alpha)+\left(
256+(1+\alpha)(
 4+\alpha-5\gamma)^4\,e^2 \right)^2, \ee
and%
\be %
  q=\frac{Q}{\frac{\pi}{2}\sqrt{\lambda}\,T_{\mathcal{R}^2}^2}.
\ee %
Regarding to small values of $\alpha$ and $\gamma$, as it is
expected, the above Ohmic conductivity reproduces
\eqref{conductivity1} for large value of electric field at fixed
temperature. Moreover, as original case introduced in \cite{Karch},
although two types of charge carries contribute to the conductivity,
which are charge carries represented by $q$ and charge carries
thermally produced, the value of them are corrected by
${\mathcal{R}^2}$ correction. For example as it was pointed out in
\eqref{sigma2}, in the case of massive charge carriers
$\hat{\sigma}$ dominates in \eqref{R2cond}. Based on the small
values of $\alpha$ and $\gamma$, we expand this term and keep only first order terms as%
\be
(\hat{\sigma}_{\mathcal{R}^2})^2=(\hat{\sigma})^2\left(1+\left(\frac{1-e^2}{1+e^2}\right)\alpha+\frac{1+5e^2+5e^4}{(1+e^2)^2}\gamma\right),
\ee%
where%
\be (\hat{\sigma})^2=\frac{q}{1+e^2}.\ee%

In five dimensions, the most general theory of gravity with
quadratic powers of curvature is GB theory. The exact solutions and
thermodynamic properties of the black brane in GB gravity were
discussed in \cite{Cai:2001dz,Nojiri:2001aj,Nojiri:2002qn}. We are
going to more understand about the DC conductivity in the GB
gravity. The black brane solution in this geometry is given by
\cite{Cai:2001dz}
\begin{equation}
ds^2=-a \,\frac{u^2}{R^2}\, h(u)\, dt^2+\frac{du^2}{\frac{u^2}{R^2}
h(u)}+\frac{u^2}{R^2} \,d\vec{x}^2\label{GBmetric},
\end{equation}
where
\begin{equation}
h(u)= \frac{1}{2\lambda_{GB}}\left[ 1-\sqrt{1-4 \lambda_{GB}\left(
1-\frac{u_0^4}{u^4} \right)}\right].
\end{equation}
In (\ref{GBmetric}), $a$ is an arbitrary constant which specifies
the speed of light of the boundary gauge theory and we choose it to
be unity. As a result at the boundary, where $u\rightarrow\infty$,
\begin{equation}
 h(u)\rightarrow \frac{1}{a }, \,\,\,\,\, a= \frac{1}{2}\left(
 1+\sqrt{1-4 \lambda_{GB}} \right)\label{a}.
\end{equation}
We assume $\lambda_{GB}\leq\frac{1}{4}$, the reason is that beyond
this point there is no vacuum AdS solution and one cannot have a
conformal field theory at the boundary. The temperature is given by
\begin{equation}
 T_{GB}=\frac{\sqrt{a}\,\, u_0}{\pi \,}.
\end{equation}
In this case \eqref{1} and (\ref{Ohmic}) lead to %
\be %
 w^4\left( 1-\sqrt{1-4\lambda_{GB}(1-\frac{1}{w})}
 \right)-2ae^2\lambda_{GB}=0,
\ee %
and %
\be\label{GBconduc}\begin{split} %
 (\sigma^{GB}_{xx})^2&=\frac{4\,q^2}{2(1+ae^2)+\sqrt{2g\left(\lambda_{GB},e\right)}}\cr
 &+\frac{1}{32\pi^2\,a}
 N_c^2N_f^2\,T^2_{GB} \cos^6{\theta_*}
 \left(2(1+ae^2)+\sqrt{2g\left(\lambda_{GB},e\right)}\right)^{\frac{1}{2}},
\end{split}\ee %
where%
\be %
 g(\lambda_{GB},e)=2(1+2ae^2)+2e^4(a-\lambda_{GB})(1+\lambda_{GB}),
\ee %
and%
\be
 e=\frac{E}{\frac{\pi}{2}\sqrt{\lambda}\,T_{GB}^2},\,\,q=\frac{Q}{\frac{\pi}{2}\sqrt{\lambda}\,T_{GB}^2}.
\ee It is clearly seen that by setting $\lambda_{GB}=0$, one finds
the Ohmic conductivity in AdS-Schwarzschild background.
Interpretation of two terms contributing to Ohmic conductivity in
\eqref{GBconduc} is the same as ${\mathcal{R}^2}$ correction case.

\section{Discussion and Conclusion}

In this paper effect of curvature correction on DC conductivity is
studied. In particular we considered ${\cal{R}}^4$ correction to
AdS-Schwarzschild black brane. Our results show that there is a
critical value for electric field in which correction to Ohmic
conductivity vanishes. Moreover it was shown that, depending on
other parameters, Ohmic conductivity can increase or decrease. In
another case, when magnetic field is turned on, Hall and Ohmic
conductivities are plotted in Fig. 1-4. At fixed temperature and
large magnetic field, Ohmic and Hall conductivity generally start
from their maximum value. Fig. 4 shows that Hall conductivity has a
maximum value at fixed electric field and temperature.

Curvature-squared and Gauss-Bonnet corrections are then considered
in the next section. As it is expected, the main structure of Ohmic
conductivity is preserved by corrections.

It is straightforward to calculate the DC conductivity from the drag
force calculations. One should consider massive carriers as endpoint
of strings which end on flavor branes and using the quasi-particle
description write the equation of motion for them at the equilibrium
where the external force is $f=E$. At large mass limit, only charge
carriers contribute to current and one may express $\langle J^x
\rangle$ in terms of velocity of the quasi-particles, $\langle J^x
\rangle=\langle J^t\rangle\,v $. Regarding Ohm's law, we have
$v=\frac{\sigma\,E}{\langle J^t \rangle}$, where $v$ is velocity of
massive carrier and it is constant. One can follow this approach and
derive $\hat{\sigma}$ in the case of $\mathcal{R}^2$ and
Gauss-Bonnet corrections in \eqref{R2cond} and \eqref{GBconduc}.
This approach has been followed in the case of strange metals in
\cite{strangeM,Fadafan:2009an}.

\section*{Acknowledgment}
M. A. would like to thank M. M. sheikh-Jabbari for instructive
discussions. We thank the organizers of ISS2010 for providing an
excellent atmosphere for physics discussions. We would like to thank
Carlos Nunez for valuable discussions and comments.

\end{document}